\documentclass[mathleft]{an}
\usepackage{graphicx}
\usepackage{euscript}
\usepackage{amsmath}	
\usepackage{amssymb}	

\usepackage{times}   
\begin{document}

\Pagespan{001}{}
\Yearpublication{2017}%
\Yearsubmission{2017}%
\Month{XX}%
\Volume{999}%
\Issue{XX}%

\title{Statistics of magnetic field measurements in OBA stars and the evolution of their
  magnetic fields}
%
%
\author{A.~S.~Medvedev\inst{1}\fnmsep\thanks{A.~S.Medvedev: \email{a.s.medvedev@gmail.com}\newline}
 \and   A.~F.~Kholtygin\inst{2}\fnmsep\thanks{A.~F.Kholtygin: \email{afkholtygin@gmail.com}\newline}
 \and   S.~Hubrig\inst{3}  
 \and   M.~Sch\"oller\inst{4}  
 \and   S.~Fabrika\inst{1,5}    
 \and   G.~G.~Valyavin\inst{1}  
 \and   G.~A.~Chountonov\inst{1}   
 \and  Yu.~V.~Milanova\inst{2}  
 \and   O.~A.~Tsiopa\inst{6}  
 \and   V.~A.~Yakovleva\inst{2}  
}

\institute{
 Special Astrophysical Obsevatory, Nizhnii Arkhyz, Russia 
 \and 
 Saint-Petersburg University, Saint-Petersburg, Russia
 \and 
 Leibniz-Institut f$\ddot{\mathrm{u}}$r Astrophysik Potsdam (AIP), Potsdam, Germany
 \and
 European Southern Observatory, Garching, Germany
 \and
 Kazan Federal University, Kazan, Russia
 \and
 Main (Pulkovo) Astronomical Observatory, Saint-Petersburg, Russia
}

\titlerunning{evolution of magnetic field of OBA stars}
\authorrunning{A.S.Medvedev et al.}
\institute{Special Astrophysical Observatory, Nizhnii Arkhyz, Russia
\and 
Saint-Petersburg University, Russia
\and 
Leibniz-Institut f$\ddot{\mathrm{u}}$r Astrophysik Potsdam (AIP), Potsdam, Germany
\and
European Southern Observatory, Garching, Germany
 \and
 Kazan Federal University, Kazan, Russia
\and
Main (Pulkovo) Observatory, Saint-Petersburg, Russia 
}

\received{~~ ~~~ 2017}
\accepted{~~ ~~~ 2017}
\publonline{later}

\abstract{We review the measurements of magnetic fields of OBA stars. Based on these data we confirm that 
          magnetic fields are distributed according to a lognormal law with a mean $\log{B}=-0.5$ ($B$ in kG) with a standard 
          deviation $\sigma=0.5$. The shape of the magnetic field distribution is similar to that for neutron 
          stars. This finding is in favor of the hypothesis that the magnetic field of a neutron star is 
          determined mainly by the magnetic field of its predecessor, the massive OB star. Further, we model 
          the evolution of an ensemble of magnetic massive stars in the Galaxy. We use our own population 
          synthesis code to obtain the distribution of stellar radii, ages, masses, temperatures, effective 
          magnetic fields, and magnetic fluxes from the pre-main sequence via zero age main sequence 
          (ZAMS) up to the terminal age main sequence stages. A comparison of the obtained in our model 
          magnetic field distribution (MFD) with that  obtained from the recent measurements of the stellar 
          magnetic field allows us to conclude that the evolution of magnetic fields of massive stars is slow 
          if not absent. The shape of the real MFD shows no indications of the {\it magnetic desert} proposed 
          previously. Based on this finding we argue that the observed fraction of magnetic stars
          is determined by physical conditions at the PMS stage of stellar evolution.         
         }

\maketitle

\section{Introduction}

\label{s:introduction}

Knowledge of the properties of the magnetic fields in massive stars is very important for our understanding the mechanisms of 
their formation and evolution as well as their impact on the stellar parameters and evolution. Accurate studies of
the age, environment, and kinematic characteristics of magnetic stars are promising to give us new insight 
into the origin of the magnetic fields~(e.g. \cite{Hubrig-2011}, 2013; \cite{Gonzalez-2017}).

For a number of early B-type stars the magnetic fields were detected several tens of years ago (e.g. \cite{Babcock:1947gf}). The first magnetic
field detection in an O-type star was made in 2002 by \cite{Donati-2002}, even though the existence of magnetic O-type stars
had been suspected for a long time.

The recent systematic surveys MiMeS (The Magnetism in Massive Stars) and BOB (The B fields in OB stars) aiming at the detection and studying of the magnetic fields in 
massive stars (\cite{Wade:2016es}; \cite{Morel-2014}) strongly enhanced the number of known magnetic OBA stars in 
comparison with $\sim$500 OBA stars with confirmed magnetic field which were known before 2009 
(e.g.~\cite{Bychkov:2009jj}). Because of the increase in the number of new detections it was also found that incidence of massive magnetic stars is about of 7\% only (\cite{Wade:2014cj}).

The detection rate obtained by the BOB collaboration is of $6\pm 4\%$ (\cite{Scholler-2017}) which is consistent 
with that given by the MiMeS group.

Measuring the magnetic fields of hundreds of OBA stars opens the possibility to study the magnetic field 
distribution for different types of stars~(e.g., \cite{Kholtygin-2010a}). The understanding of the nature 
of the magnetic field of massive stars can be based on our knowledge of their MFD. Checking the real MFD 
showed that some previous ideas about the magnetic field evolution can be incorrect. 

For example, the MFD predicted by \cite{Ferrario-2006} for massive stars on the main sequence in the mass range
from 8 to 45\,$M_{\odot}$ 
based on the hypothesis that the net magnetic flux for a massive star and its
descendant, the neutron star, is the same (see their Fig.~4) does not agree with the real distribution of the magnetic 
field (e.g.\ \cite{Kholtygin-2010a}). 

The disagreement between the predicted MFD and the one based on the real
measurements is connected with the dissipation of 
the stellar magnetic flux during the evolution from massive stars to neutron stars 
(see Fig.~1 in the paper by~\cite{IgoshevKholtygin2011} and subsection~\ref{s:magnetic_flux} in the present paper). 
It is shown that the magnetic fluxes of neutron stars are about of three orders of magnitude lower than for their 
progenitors, the massive OB stars.

It means that the studying how the magnetic fields and the magnetic fluxes change during the massive star 
evolution plays a decisive role in the understanding the nature of the magnetic field of this group of stars. 
In the present paper we consider the evolution of the magnetic fields and fluxes of OBA stars at the main 
sequence stage. 

Our paper is organized in the following way: The accepted model of the evolution of magnetic OBA stars is described 
in Sect.~\ref{s:model}. In Sect.~\ref{s:empirical_distributions}  we review the empirical 
magnetic field distributions of OBA stars. In Sect.~\ref{s:discussion} we discuss the {\it magnetic desert} 
problem and give some constrains on the rate of dissipation of stellar magnetic fields. 
Finally we summarize our results and give some conclusions in Sect.~\ref{s.Concl}.

\section{The Model}
\label{s:model}

\subsection{Population synthesis}
\label{s:population_synthesis}

Information about the intrinsic distribution of stellar magnetic fields can only be obtained from an 
analysis of observations of certain sample of stars. 
Even the stars of the same spectral class and even subclass have significant diversity in 
masses, ages and  other characteristics. These characteristics influence the magnitude of a 
large-scale magnetic field due, for example, the geometrical effects and the dissipation processes. 
Therefore, to simulate the magnetic field distribution for the model ensemble of stars, these variations in the
stellar parameters have to be taken into account. Here we outline the main features of our population 
synthesis model realizing this aim.

The first step of our modeling is a creation of the initial ensemble of stars. We suppose that a total number 
of stars in the ensemble is $N_\mathrm{tot}$ and the stellar masses 
$M \in \left[M_\mathrm{min},M_\mathrm{max}\right]$. The mass distribution at the ZAMS is descibed by a power 
law with an exponent of $-2.3$. This corresponds to the high-mass region of the stellar initial mass 
function (\cite{Kroupa:2002cm}). The appearance time $t_*$ of an each star at ZAMS is generated randomly 
using the uniform  distribution in the range $t_* \in \left[0,T\right]$, where $T$ 
is the total simulation time. We use the constant stellar birthrate $\lambda$. 

This means that approximately one star appears in the time interval $\Delta T=1/\lambda$. The simulation 
time $T$ has to be at least three times longer than the main sequence lifetime $\tau_\mathrm{MS}$ of a least 
massive star in the ensemble to be sure that the model star ensemble is stationary.

Next steps include the computation of stellar parameters and selection of the stars remaining on the main 
sequence at the moment of time $T$. Evolution of the stars in the ensemble is simulated using the rapid 
single-star evolution code \textsc{SSE} developed by \cite{Hurley:2000gx}. This code is based on analytical 
approximations of evolutionary tracks computed by \cite{Pols:1998fk}. It can be used for simulations of 
all stages of stellar evolution from the ZAMS to final remnants and it is valid for masses in the range 
$0.1$--$100$~$M_{\odot}$.

Our population synthesis code is created on the base of the Astrophysical Multipurpose Software Environment (\textsc{AMUSE}) 
developed by \cite{Pelupessy:2013kc}. The platform \textsc{AMUSE} uses \textit{python} 
as an interface between different existing astrophysical co\-des (stellar dynamics, stellar evolution, 
hydrodynamics, etc.) and provides a framework in which these codes can be 
coupled\footnote{see {http://www.amusecode.org for details}}.

\subsection{Evolution of stellar magnetic fields}
\label{s:evolution_of_magnetic_fields}

\subsubsection{Basic definitions}
\label{s:basic_definitions}

In our model the magnetic field of a star is defined via the net magnetic flux $\Phi$ at the stellar surface:
\begin{equation}
	\Phi = \int \limits_{S} |B_r| dS.
\end{equation}

Here $B_r$ is the radial projection of the $B$-field and $dS$ is a surface element. This parametrization is 
very useful for describing the magnetic properties of stellar populations because the assumption about of the 
conservation of magnetic flux in the  absence of dynamo or dissipation mechanisms  is very good  
(e.g. \cite{Braithwaite:2006da}).

In the case of a dipole field the relation between the net magnetic flux $\Phi_d$ and the polar field 
$B_{\mathrm{d}}$ can be easily derived:
\begin{equation}
	\Phi_d = 4/3 \pi B_{\mathrm{d}} R^2,
\end{equation}
where $R$ is a radius of a star.
However, current techniques used to measure stellar magnetic fields can give us only the longitudinal component 
$B_l$ of the field. Although $B_l$ is directly connected to the polar field $B_{\mathrm{d}}$, it also depends 
on the rotation phase $\phi$, angle between the magnetic dipole and rotation axes $\beta$, and the inclination 
of the rotation axis to the line of sight $i$ (e.g. \cite{Preston:1967jr}). The later two parameters are 
completely random and have a wide range of possible values. Even the phase-averaged field 
$\langle{B_l}\rangle_\phi$ is very sensitive to their variations.

Because of these reasons, we use the root-mean-square (\emph{rms}) field $\mathcal{B}$ instead of 
$B_{\mathrm{d}}$ as the main characteristic of stellar magnetic fields. If $N$ is the total number of the 
field measurements $B_l^k$ ($k=1 \dots N$), then the \emph{rms}-field is given by the formula 
(e.g. \cite{Bohlender:1993vy}):
\begin{equation}\label{e:rms-field}
	\mathcal{B} = \sqrt{ \frac1{N} \sum\limits_{k = 1}^n \left(B_l^k\right)^2 }.
\end{equation}
The phase-averaged ratio $\mathcal{B} / B_{\mathrm{d}}$ and its asymptotical behavior at $N \to \infty$ were 
investigated by Kholtygin et al. (2010a). They demonstrated that in the case of a dipole configuration the 
\emph{rms}-field $\mathcal{B}$ weakly depends on random values of the rotational phase $\phi$, inclination 
$i$ and the angle $\beta$. This conclusion also is valid for quadrupole and more complex field configurations. 
Following this paper we adopt that
\begin{equation}
\label{e:Brms-Bd}
		\mathcal{B} \approx 0.2 B_{\mathrm{d}}.
\end{equation}
Then we can find the net magnetic flux using next relation:
\begin{equation}
\label{e:magnetic_flux}
		\Phi = 4 \pi R^2 \mathcal{B}.
\end{equation}
It gives a good estimation of the net magnetic flux for any field configurations 
(\cite{Kholtygin-2010a}) .

\subsubsection{Magnetic field function at ZAMS}
\label{s:MFF_at_ZAMS}

In order to simulate the evolution of stellar magnetic fields with the stellar age $t$ it is necessary first 
to define the initial magnetic field function for the stars at the ZAMS ($t=0$). We assume the lognormal 
distribution of the net magnetic fluxes:
\begin{equation}\label{e:MFF_at_ZAMS}
	f(\Phi \mid t = 0) = \frac{A}{\Phi \sigma} \exp
\left\{ -\frac12\left(\frac{\log\Phi - \langle \log\Phi\rangle}{\sigma} \right)^2 \right\}, 
\end{equation}
where $\Phi$ is the net magnetic flux, $\langle \log\Phi\rangle$ is the mean value of the $\log\Phi$, 
$\sigma$ is the width (in dex) of the distribution and
\begin{equation}
	A = \frac1{\sqrt{2 \pi} \ln{10}}.
\end{equation}
We chose the lognormal distribution of the magnetic fluxes mainly due to similarity between lognormal 
and empirical distributions derived from real samples of magnetic stars (this similarity is discussed in 
Section~\ref{s:empirical_distributions}). A lognormal magnetic flux distribution (magnetic flux function) 
may be obtained from the simple assumption that during the pre-main sequence evolution the magnetic field of 
a star is experiencing a finite number of cycles of amplification/damping by some random factor. Under rather 
wide conditions the resulting magnetic field function will coincide to a lognormal distribution even for 
a small number of the cycles (see \cite{Kholtygin-2016} for details).

\subsubsection{Dissipation of magnetic fields}
\label{s:dissipation_of_magnetic_fields}

Observational evidences imply that the magnetic fields of Ap stars decay. The rate of 
dissipation depends strongly on stellar mass (e.g. \cite{Landstreet:2008bi}). 
According to \cite{Kholtygin-2010b} the dissipation of magnetic fields can be represented as the 
exponential function $\propto e^{ - \alpha_{\mathrm{d}} \tau}$, where $\tau$ is the age of a star expressed 
in terms of the main sequence lifetime of the star, and $\alpha_{\mathrm{d}} \sim 2.0$ is the dissipation 
factor. We also assume exponential decay of stellar magnetic fields on the time-scale $t_{\mathrm{d}}$ which 
is set to be proportional to the lifetime of a star on the main sequence $t_{\mathrm{MS}}$. Introducing the 
dissipation parameter as
\begin{equation} 
\label{e:dissipation_parameter}
	\tau_{\mathrm{d}} = \frac{ t_{\mathrm{d}} }{ t_{\mathrm{MS}} },
\end{equation}
we obtain the following expression for the temporal evolution of the net magnetic flux:
\begin{equation}
	\Phi(t) = \Phi(0) \exp\left\{-\frac{1}{\tau_\mathrm{d}} 
         \left(\frac{t}{t_\mathrm{MS}} \right) \right\}.
\end{equation}
The time dependence of the \emph{rms}-field $\mathrm{B}(t)$ can be derived from this relation and by
using the formula (\ref{e:magnetic_flux}).

\subsection{Random sampling and parameter estimation}
\subsubsection{Generation of empirical distributions}
\label{s:random_sampling}

To derive an empirical magnetic fields distribution from a sample consisting of $N_\mathrm{smpl}$ 
stars, the first step is to split the entire range of magnetic fields into a set of $N_{\mathrm{bins}}$ 
equally-sized intervals (bins) and then count how many stars fall into each bin. Obviously, large samples are 
more preferable for obtaining empirical distributions which would accurately resemble the intrinsic magnetic 
field function. Unfortunately, the current number of known magnetic stars is still very limited, 
therefore it is important to understand 
i) how the {\it intrinsic} model magnetic field function would manifest itself as empirical distribution and 
ii) how the number of stars in a sample affects the magnitude of possible variations in distribution function 
due to statistical effects. For these purposes we developed the random sampling method which allows to 
produce empirical distributions from the model.

The random sampling is used to generate a large set of equally-sized samples from the ensemble of magnetic 
stars created by the population synthesis module (see Section~\ref{s:population_synthesis}). Then 
distributions of magnetic fields in each of the samples are passing through the special procedure to obtain 
average distribution as well as confidence boundaries for the uncertainties in a counts number for each of 
the defined bins.

If $n_k(i)$ is the number of stars in $i$-th bin of $k$-th sample then the averaged 
number of stars in the bin is 
\begin{equation}
	\langle{n(i)}\rangle = \frac1{N_\mathrm{smpl}} \sum\limits_{k=1}^{N_\mathrm{smpl}} n_k(i),
\end{equation}
where $N_\mathrm{smpl}$ is the number of generated samples. The confidence boundaries for a desired level of 
significance are calculated using the Poisson statistics.

\subsubsection{Parameter estimation}
\label{s:parameter_estimation}

Since real samples of massive magnetic stars are small, the standard $\chi^2$ statistics is no longer valid 
for the parameter estimation. For this reason, instead of $\chi^2$ we use the $C$ statistics in form 
%
\begin{equation}
	C_q = 2 \sum\limits_{i=1}^{N_\mathrm{{bins}}}
	\left[ e_i - n_i + n_i ( \ln{n_i} - \ln{e_i} ) \right],
\end{equation}
where $n_i$ is a number of stars in each of the bins of the empirical distribution, $e_i$ is the 
expected number of stars given by the model, and $N_\mathrm{bins}$ is a number of bins. The $C$ statistics were introduced by \cite{Cash:1979bi} as replacement of $\chi^2$ 
that would be valid even for a very small samples. It is commonly used in X-ray and 
gamma-ray astronomy in cases when the number of photons received by a detector is low.

The following method is used for generating confidence intervals for the model parameters. First, we find the 
minimum value $(C_q)_\mathrm{min}$ when the model is fitted to the data and $q$ parameters are varied. Then, 
assuming the level of significance $\alpha$, we determine the locus of points in parameter space for which
\begin{equation}
		C - (C_q)_\mathrm{min} > \Delta C(\alpha).
\end{equation}
According to \cite{Cash:1979bi} the $\Delta C$ statistics can be represented as
\begin{equation}
		\Delta C(\alpha) = \chi^2_q(\alpha) + O\left( \frac1{\sqrt{N_\mathrm{smpl}}} \right),
\end{equation}
where $N_\mathrm{smpl}$ is the total number of stars in a samples. For $N_\mathrm{smpl} \gg 1$ the $\Delta C$ statistics follows 
the $\chi^2$ distribution for $q$ degrees of freedom. However, we generate the distribution of $\Delta C$ 
from our model using the Monte Carlo simulations, because the term $O(N_\mathrm{smpl}^{-1/2})$ cannot be neglected 
when $N_\mathrm{smpl} \lesssim 10$.

The $C$ statistics may also be easily modified for the purpose of simultaneous fitting of different datasets.

\subsection{Parameters of the model}
\label{s:model_parameters}

\begin{table}
\caption{List of the main model parameters.}
\label{t:model_parameters}
 \begin{tabular}{lccl}
\#  &  Quantity                     &  Units          &  Description         \\   \hline
 1. &$M_\mathrm{min},M_\mathrm{min}$&$M_{\odot}$      & Mass interval limits \\
 2. & $N_\mathrm{smpl}$             &                 & Number of stars      \\
    &                               &                 & in a sample          \\
 3. &$\langle\log\Phi\rangle$       &G$\cdot$cm$^2$   & Mean magnetic flux   \\
    &                               &                 & at the ZAMS          \\
 4. & $\sigma$                      & dex             & Width of the initial \\
    &                               &                 & distribution         \\
 5. & $\tau_\mathrm{d}$             &$t_\mathrm{MS}^a$& Time-scale of the    \\
    &                               &                 & magnetic field       \\   \hline
 \end{tabular}
\flushleft $^{a}$ Lifetime of a star on the main sequence.
\end{table}

A short summary describing the main parameters of the model is presented in Table~\ref{t:model_parameters}. 
Note that two of the parameters such as the mass interval $[M_\mathrm{min}, M_\mathrm{max}]$ and the sample 
size $N_\mathrm{smpl}$, may be determined directly from characteristics of empirical data. For example, an 
empirical magnetic fields distribution derived from a sample of eleven O-type stars suggests that 
$M \in [16, 60]\,M_{\odot}$ and $N_\mathrm{smpl} = 11$. The remaining three parameters however have to be 
estimated from fitting of the empirical data.

 \begin{figure}
 \centering
 \includegraphics[width=0.95\linewidth]{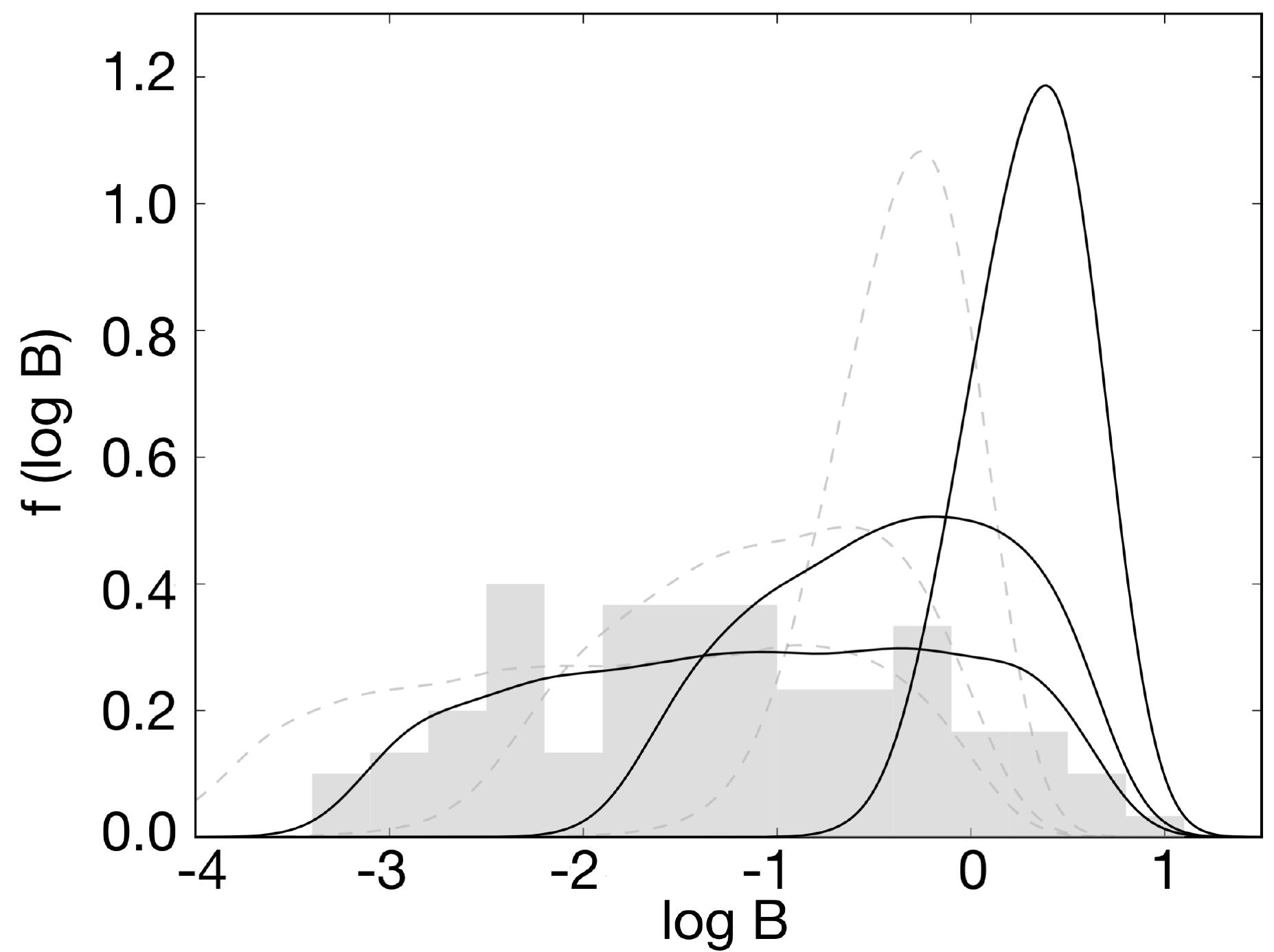}
 \caption{Magnetic field function of early-type stars calculated using our model. Black lines correspond to 
          different values of the dissipation parameter $\tau_{\mathrm{d}}=0.15$ (left curve), $0.3$ (central 
          curve), and $\infty$ (right curve) in Eq.\ref{e:dissipation_parameter}. All distributions are obtained
          from the stellar ensemble containing stars with masses in the range $2$--$10$~$M_{\odot}$ and 
          following parameters of the net magnetic flux distribution at the ZAMS: 
          $\langle\log\Phi\rangle = 27.5$, $\sigma = 0.2$ (see Eq.\ref{e:MFF_at_ZAMS} for detail). 
          Dashed lines show the distributions calculated using the same parameters but for the mass 
          range $10$--$60$~$M_{\odot}$. 
          The gray histogram represents a single realization of the empirical distribution 
          ($\tau_{\mathrm{d}} = 0.15$) for the small ensemble with $N_\mathrm{smpl} = 100$ 
          for the initial masses in the $2$--$10$~$M_{\odot}$ range. The magnetic field strength $B$ is given in kG.
          }
\label{f:model}
 \end{figure}

Some examples of magnetic field functions produced by the model are presented in Fig.~\ref{f:model}. 
Such parameters as $\langle\log\Phi\rangle$ and $\sigma$ define the position and overall width of 
the magnetic fields function. The mass interval influences some minor asymmetry in the shape of the magnetic 
field function due to changes of stellar radii during evolution on the main sequence. The dissipation 
parameter $\tau_\mathrm{d}$ has the most profound impact on a magnetic field function. 
In the case of no dissipation of magnetic fields, or if $\tau_\mathrm{d} \gg 1$, the magnetic field function 
would  be a bell-shaped curve, with some minor asymmetry. However, if the magnetic field decay is fast 
($\tau_\mathrm{d} \ll 1$),  the model would produce a highly asymmetrical magnetic field function. Finally, 
the sample size parameter $N_\mathrm{smpl}$ defines possible manifestations of a magnetic field function as 
an empirical distribution derived form a finite number of magnetic fields measurements.

\section{Analysis of the empirical magnetic field distributions for OBA stars}
\label{s:empirical_distributions}

\subsection{Magnetic field measurements}
\label{s:magnetic_field_measurements}

In order to compare our model with observations, we first divided all known magnetic stars into three groups: 
BA stars (this groups contains mainly Ap/Bp stars), O and early B type stars combined together (OB stars), 
and O type stars only. These groups correspond to the mass ranges $1.5$--$16$, $3$--$16$ and 
$16$--$60\,M_{\odot}$.

Magnetic field measurements for BA stars were taken from the catalogue by Bychkov et al. (2009) which contains 
hundreds of magnetic stars. However only $288$ stars with most reliable measurements were selected, based
on statistical criteria described by Kholtygin et al. (2011a). The data for O- and early B-type stars were 
compiled from multiple sources 
(\cite{Petit:2013kp}; \cite{Fossati:2014ce}, 2015a, 2015b; \cite{Alecian:2014cr}; \cite{Castro:2015bv}) with the 
total amount of $73$ stars, including $11$ O-type stars.

If only dipolar magnetic field strengths $B_\mathrm{d}$ were given in the used sources, we converted them to 
\emph{rms} magnetic field $\mathcal{B}$ via relation~(\ref{e:Brms-Bd}).
The empirical magnetic field distribution $f(\mathrm{B})$ for the sample of stars with measured magnetic 
fields was introduced by \cite{Fabrika-1999} and \cite{Monin-2002}. It equals to 
\begin{equation}
\label{Eq.MFFdeterm}
f(\mathcal{B})  \approx \frac{N(\mathcal{B},\mathcal{B} +\Delta \mathcal{B})}
         {N_{\mathrm{tot}}\Delta \mathcal{B}} \, , 
\end{equation}
where $N(\mathcal{B},\mathcal{B}+\Delta\mathcal{B})$ is the number of stars with the \emph{rms} magnetic 
fields $\mathcal{B}$ in the interval $(\mathcal{B}, \mathcal{B}+\Delta \mathcal{B})$ and $N_{\mathrm{tot}}$ 
is the total 
number of stars with measured {\it rms} magnetic field.

\subsection{BA stars}
\label{s:Ap/Bp}

\begin{figure}
\centering
\includegraphics[width=0.95\linewidth]{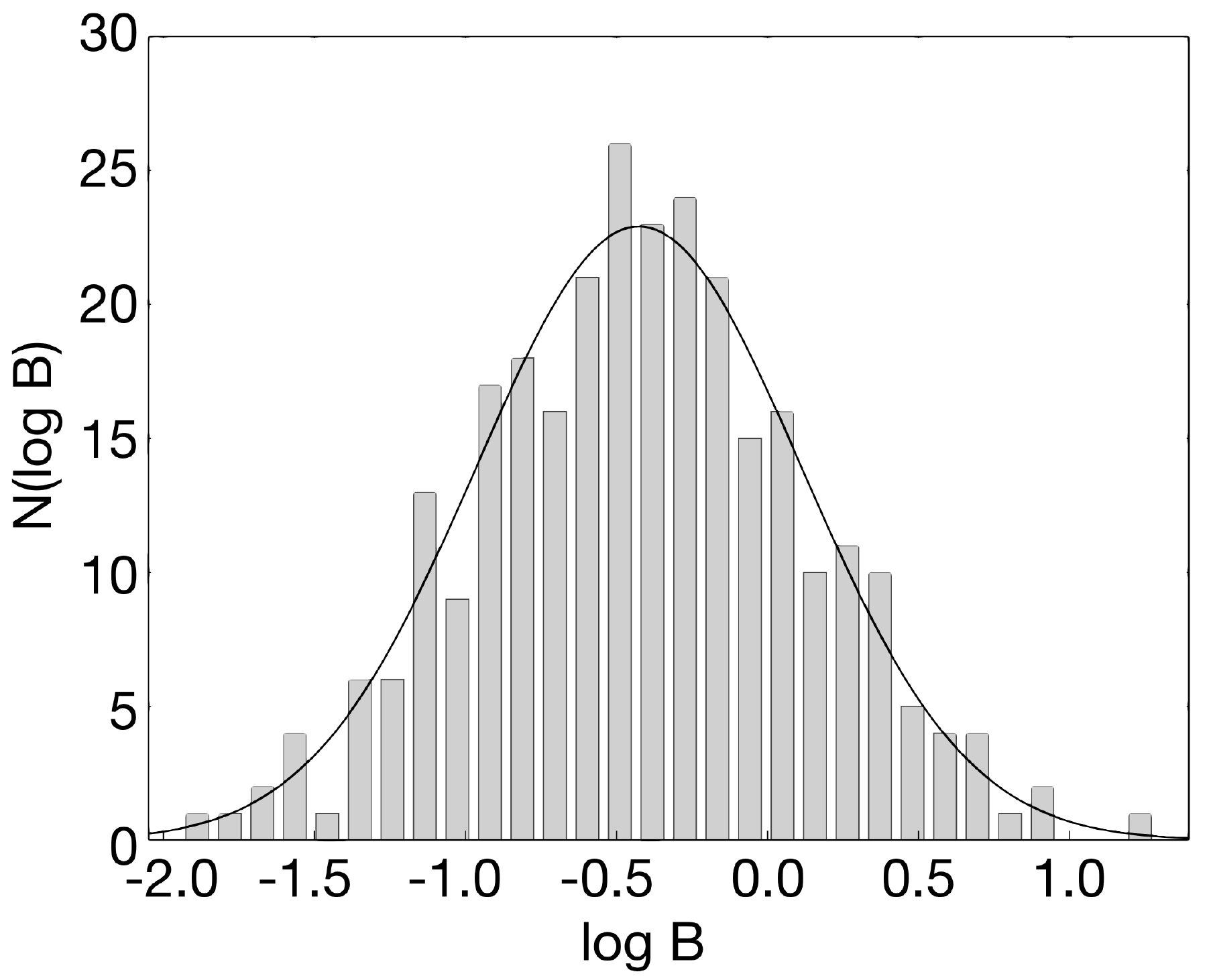}
\caption{Magnetic field distribution for BA stars (gray histogram) $N=N(\mathcal{B},\mathcal{B}+\Delta\mathcal{B})$.
         The distribution is derived using 
         the measured magnetic field data by Bychkov et al. (2009). Black line represents the 
         lognormal distribution with $\langle\log{B}\rangle \approx -0.5$ ($B$ in kG) and $\sigma \approx 0.5\,$dex.
}
\label{f:bychkov0}
\end{figure}

The empirical distribution of magnetic fields derived from our sample of BA stars is presented in 
Fig.~\ref{f:bychkov0}. It is bell-shaped with a maximum located at $\mathcal{B} \approx 300\,$G, and may be 
easily approximated by lognormal distribution. In Section~\ref{s:model_parameters} we found that the shape of 
the model distribution was highly dependent on the value of the dissipation parameter (see also 
Fig.~\ref{f:model}). Therefore, one may expect that $\tau_\mathrm{d}$ has to be large to produce the 
symmetrical magnetic field functions which would be consistent with the empirical data  in 
Fig.~\ref{f:bychkov0}. At the same time the maximal symmetry would be achieved 
only if the dissipation parameter was infinite. 
Thus we decided to consider two models: 
i) the model $\mathcal{M}_0$ in which we assume no dissipation of stellar magnetic fields, i.e. 
  $\tau_\mathrm{d} \to \infty$; 
ii) the model $\mathcal{M}_1$ with a dissipation and a finite value of $\tau_\mathrm{d}$.

\begin{figure*}
\centering
\includegraphics[width=0.47\linewidth]{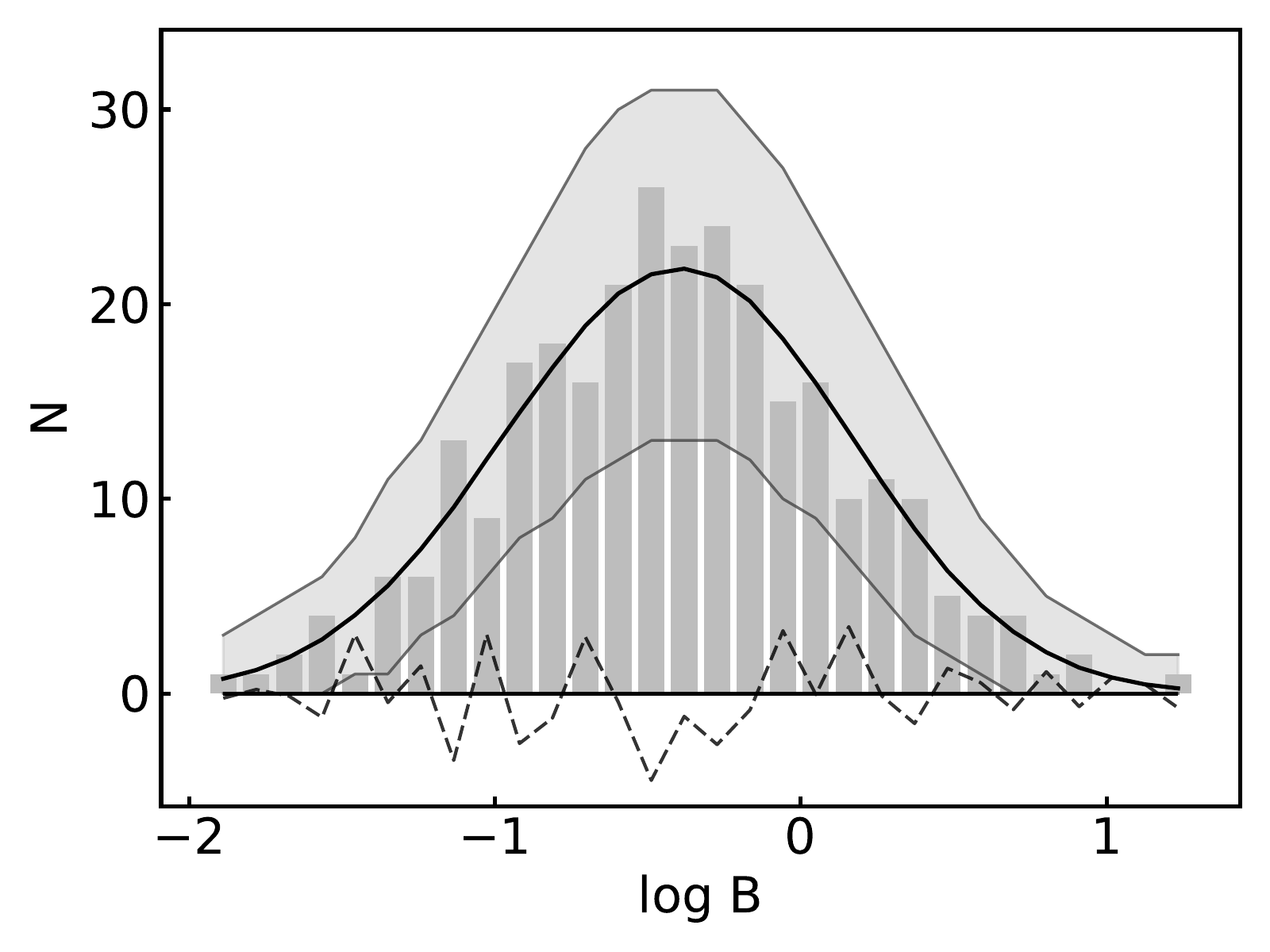}
\hspace{0.5cm}
\includegraphics[width=0.47\linewidth]{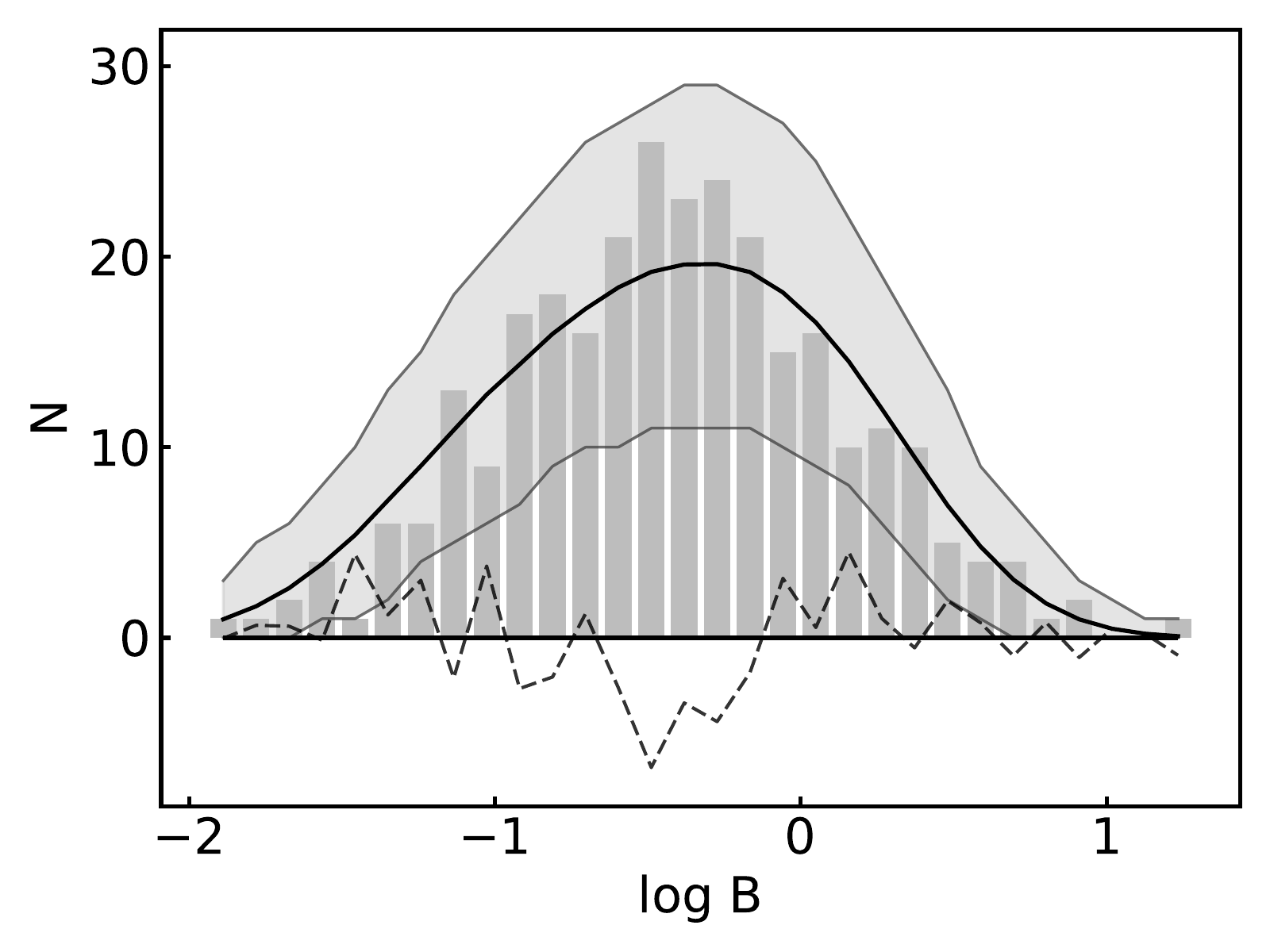}
\caption{{\it Right panel.} Model $\mathcal{M}_0$ (thick black line) in comparison with the empirical magnetic field 
             distribution for BA stars (grey historgram). The dashed line represents the residuals which are 
			 the difference between the model and the observed distribution. The model 
			 distribution is calculated for the 
             parameters providing the best fit $\langle\log\Phi\rangle= 26.42$, $\sigma = 0.5$. The dissipation 
             parameter $\tau_\mathrm{d}=\infty$. The gray filled area corresponds to the 95\,\% confidence 
             interval for the possible variations of the sampled magnetic field distribution with upper and lower limits represented by the thin grey lines.
			 {\it Left panel.} The same but for the model $\mathcal{M}_1$. The parameters 
             of the best fit are $\langle\log\Phi\rangle = 26.83$ and $\sigma= 0.33$. The dissipation parameter 
             is fixed at $\tau_\mathrm{d} = 0.5$. 
}
\label{f:Ap/Bp}
\end{figure*}


\begin{table}
{\small
\renewcommand{\tabcolsep}{0.10cm} 
\caption{\small Parameters of the best fitting models for magnetic field distribution functions of BA 
                and OB stars.}
\label{t:fits}
	\begin{tabular}{lcccccc}  
		\hline
      & Model& $\langle\log\Phi\rangle$ & $\sigma_0$ & $\tau_\mathrm{d}$ & $C / \mathrm{nbins}$ & $p_\mathrm{null}$ \\ \noalign{\smallskip}
      &      & $\mathrm{G}\,\mathrm{cm}^2$ &  &  &  &  \\ \hline \noalign{\smallskip}
BA    & $\mathcal{M}_0$& $26.42^{+0.04}_{-0.04}$ & $0.5^{+0.04}_{-0.05}$ & $\infty$ & 14.84/30 & 0.96\\ \noalign{\medskip}
      & $\mathcal{M}_1$& $26.83^{+0.07}_{-0.06}$ & $0.33^{+0.04}_{-0.07}$   & 0.5   & 22.2/30  & 0.37 \\ \noalign{\bigskip}
OB    & $\mathcal{M}_0$&$26.51^{+0.11}_{-0.15}$& $0.53^{+0.07}_{-0.11}$ & $\infty$  & 13.78/18 & 0.72 \\  \noalign{\medskip}
      & $\mathcal{M}_1$& $26.9^{+0.13}_{-0.17}$ & $0.47^{+0.08}_{-0.17}$ & 0.5      & 16.5/18  & 0.3 \\ \noalign{\bigskip}
 O    & $\mathcal{M}_0$& $26.95^{+0.35}_{-0.45}$ & $0.52^{+0.18}_{-0.49}$ & $\infty$& 2.67/5   & 0.35 \\ \noalign{\medskip}
      & $\mathcal{M}_1$& $27.29^{+0.47}_{-0.59}$ & $0.51^{+0.04}_{-0.1}$ & 0.5      & 3.58/5   & 0.3 \\ \noalign{\bigskip}
\multicolumn{7}{c}{Simultaneous fitting} \\ \noalign{\bigskip}
OBA   & $\mathcal{M}_0$ &$26.45^{+0.05}_{-0.05}$ & $0.5^{+0.04}_{-0.05}$ & $\infty$ & 48.3/53  & 0.73\\ \noalign{\medskip}
      & $\mathcal{M}_1$ &$26.87^{+0.05}_{-0.07}$ & $0.35^{+0.04}_{-0.09}$  & 0.5    & 48.3/53  & 0.43 \\ \noalign{\bigskip} \hline
\end{tabular}
}
\end{table}

We find that the model $\mathcal{M}_0$ is in a good agreement with the empirical magnetic field distribution 
for BA stars (see Fig.~\ref{f:Ap/Bp}, top panel). The parameters of the best-fit and the  corresponding 
confidence intervals as well as the value of the $C$ statistics are presented in Table~\ref{t:fits}.

Applying the model $\mathcal{M}_1$ for the analysis of the empirical data we discovered that from a 
statistical point of view the results were equally good for any value of the dissipation parameter taken 
from the range $\tau_\mathrm{d} \gtrsim 0.5$. Below this threshold the quality of fitting is rapidly 
decreasing beyond acceptable limits. Thus we conclude that $\tau_\mathrm{d} = 0.5$ corresponds to the 
highest rate of magnetic field decay allowed by the agreement with the empirical distribution of BA stars. 
We assume this value for the futher analysis using $\mathcal{M}_1$ model.

The model $\mathcal{M}_1$ which provide the best fit  of the empirical distribution of BA stars magnetic 
fields is given in Fig.~\ref{f:Ap/Bp} (bottom panel). Like the model $\mathcal{M}_0$ it also gives a good 
agreement with the data but it also posess the higher and less scattered values of the net magnetic flux 
$\langle\log\Phi\rangle$ for stars at the ZAMS as it seen in Table~\ref{t:fits}.

\subsection{OB stars}
\label{s:OB}

\begin{figure}
\centering
\includegraphics[width=\linewidth]{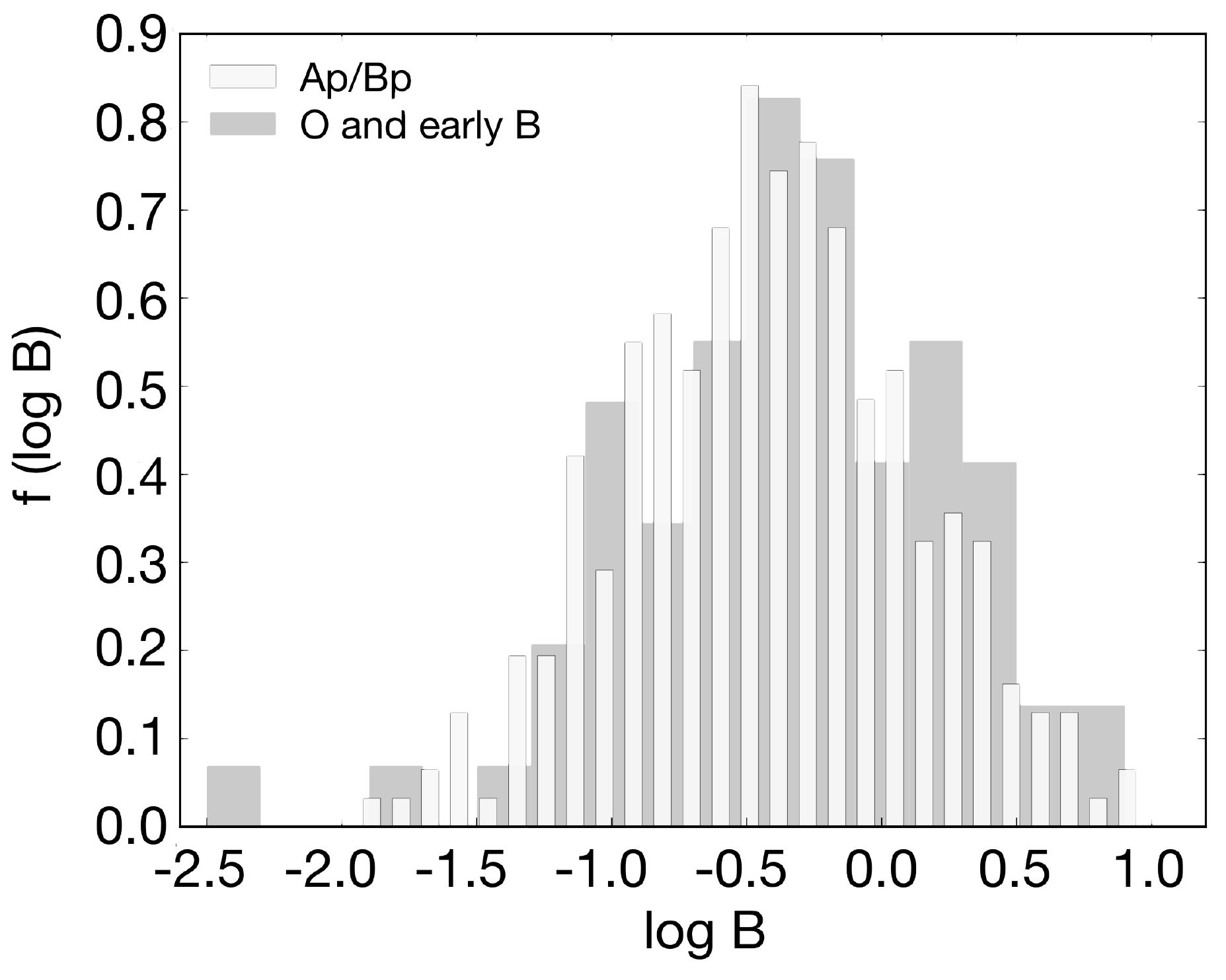}
\caption{Normalized magnetic field distributions for BA (white histogram) and OB stars 
         (gray histogram). Magnetic field strengths for BA stars were taken from Bychkov et al. (2009). 
         For OB stars data from multiple sources were used (Section~\ref{s:magnetic_field_measurements}). Both 
         distributions have the same bell-shaped appearance, location of maximum and spread.
        }
\label{f:OB}
\end{figure}    

The empirical magnetic field distribution of O- and early B-type stars is very similar in 
appearance to that of BA stars. The resemblance becomes obvious 
from Fig.~\ref{f:OB} in which these distributions are presented together. This is why first we used the 
previously obtained for BA stars best-fitting parameters to calculate the model 
distribution of OB stars. Thus, we confirmed our hypothesis that both empirical distributions are 
likely drawn from the same intrinsic magnetic field function (the corresponding $p$-values for two of the 
models are $p_\mathrm{null}(\mathcal{M}_0) = 0.4 $ and $p_\mathrm{null}(\mathcal{M}_1) = 0.31$).

\begin{figure*}
\centering
\includegraphics[width=0.47\linewidth]{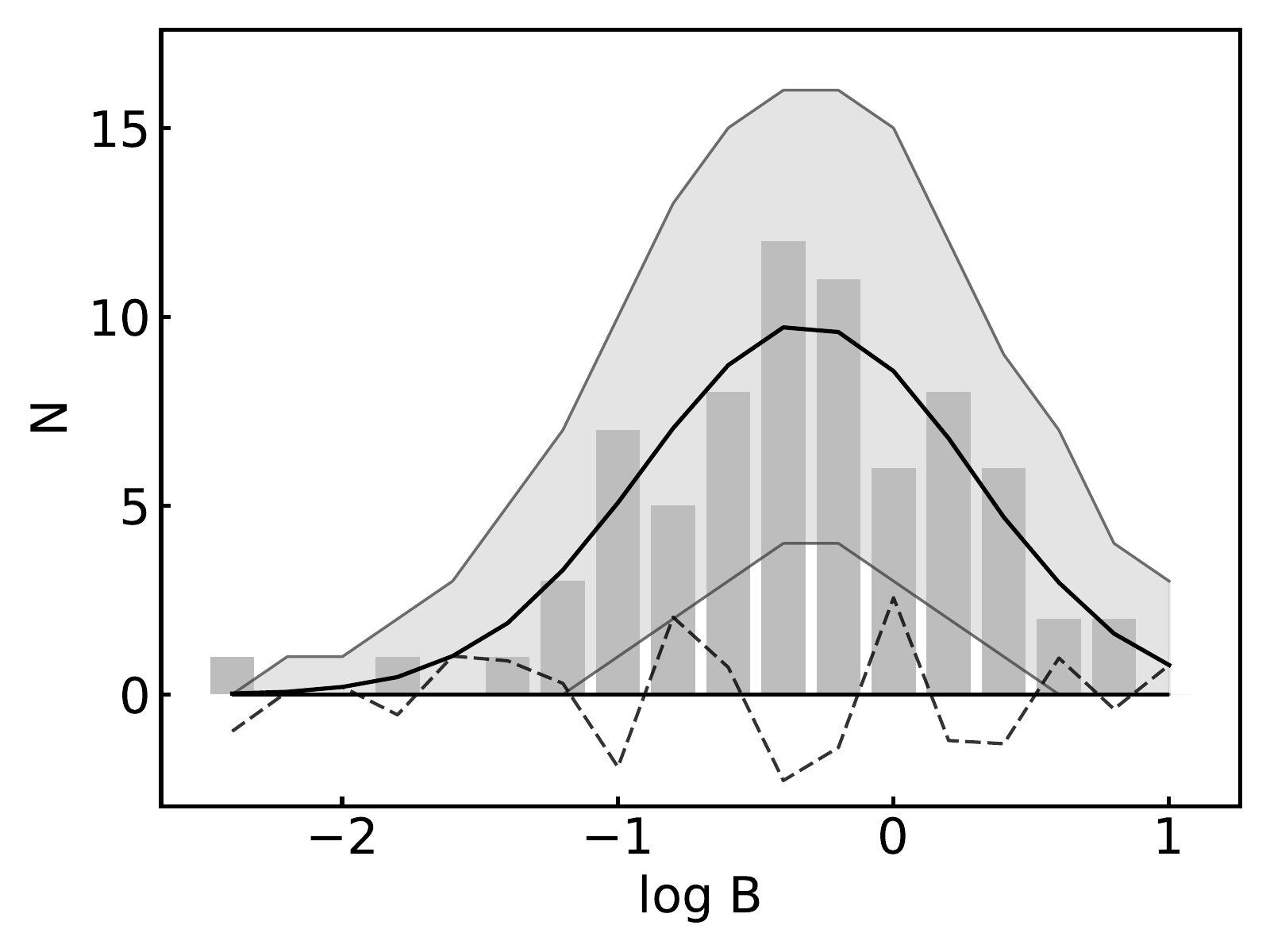}
\hspace{0.5cm}
\includegraphics[width=0.47\linewidth]{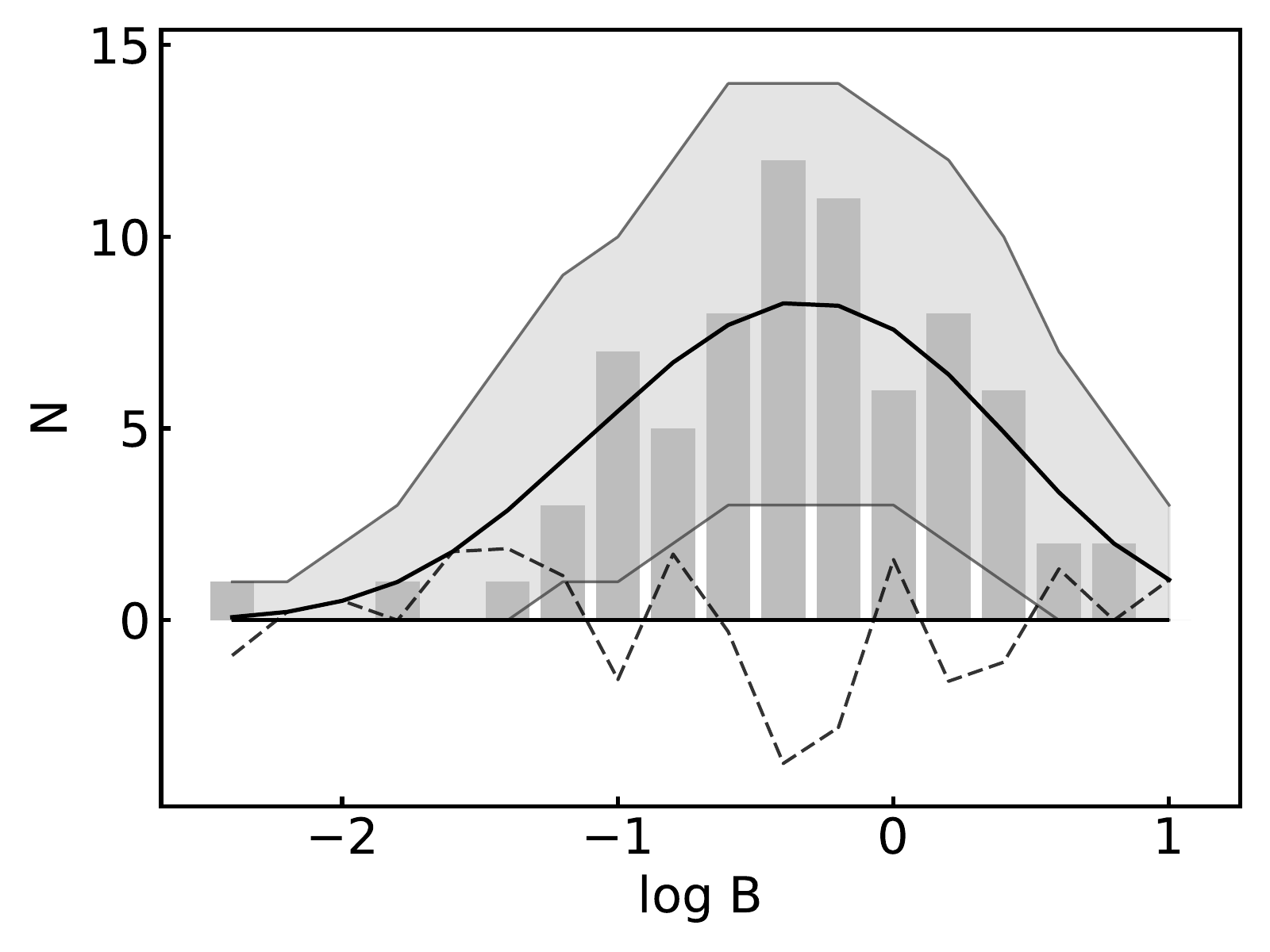}
\caption{Same as in Fig.~\ref{f:Ap/Bp}, but for O- and early B-type stars.  
         }
\label{f:OBA}
\end{figure*}

Both the model $\mathcal{M}_0$ and model $\mathcal{M}_1$ are in a good agre\-ement with the empirical 
magnetic field distribution for OB stars (see Fig.~\ref{f:OBA}). The parameters of the best-fit 
and the  corresponding confidence intervals as well as the value of the $C$ statistics are presented in 
Table~\ref{t:fits}.
We compared the proper best-fitting parameters for OB stars with those obtained 
for BA stars (Table~\ref{t:fits}). We also find that both models give a 
good agreement with the data (Fig.~\ref{f:OB}). Therefore we can conclude that the results of our 
analysis strongly imply a close similarity between the magnetic field distribution for BA and OB stars.

\subsection{O stars and OBA stars}
\label{s:O}

\begin{figure*}
\centering
\includegraphics[width=0.47\linewidth]{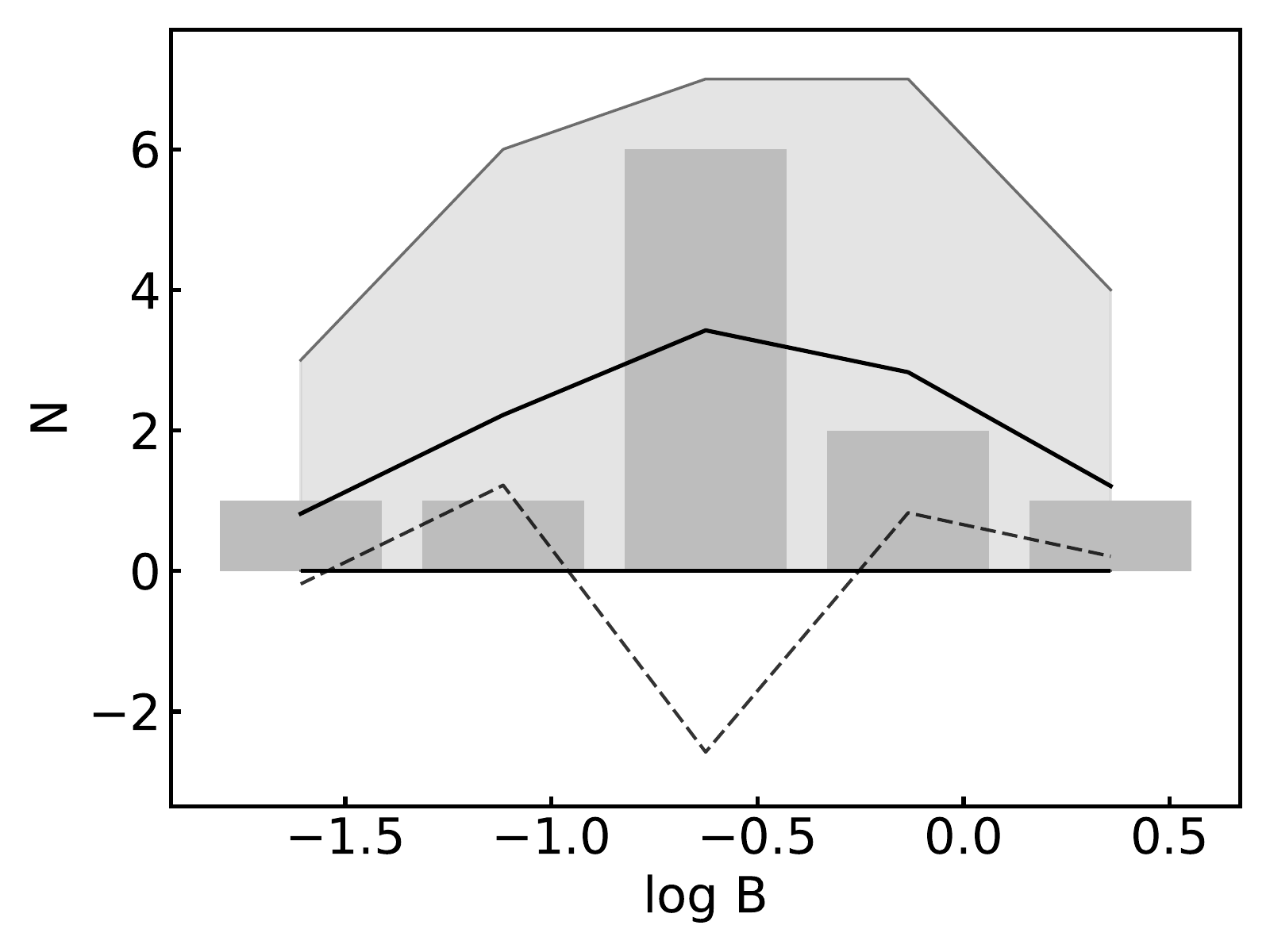}
\hspace{0.5cm}
\includegraphics[width=0.47\linewidth]{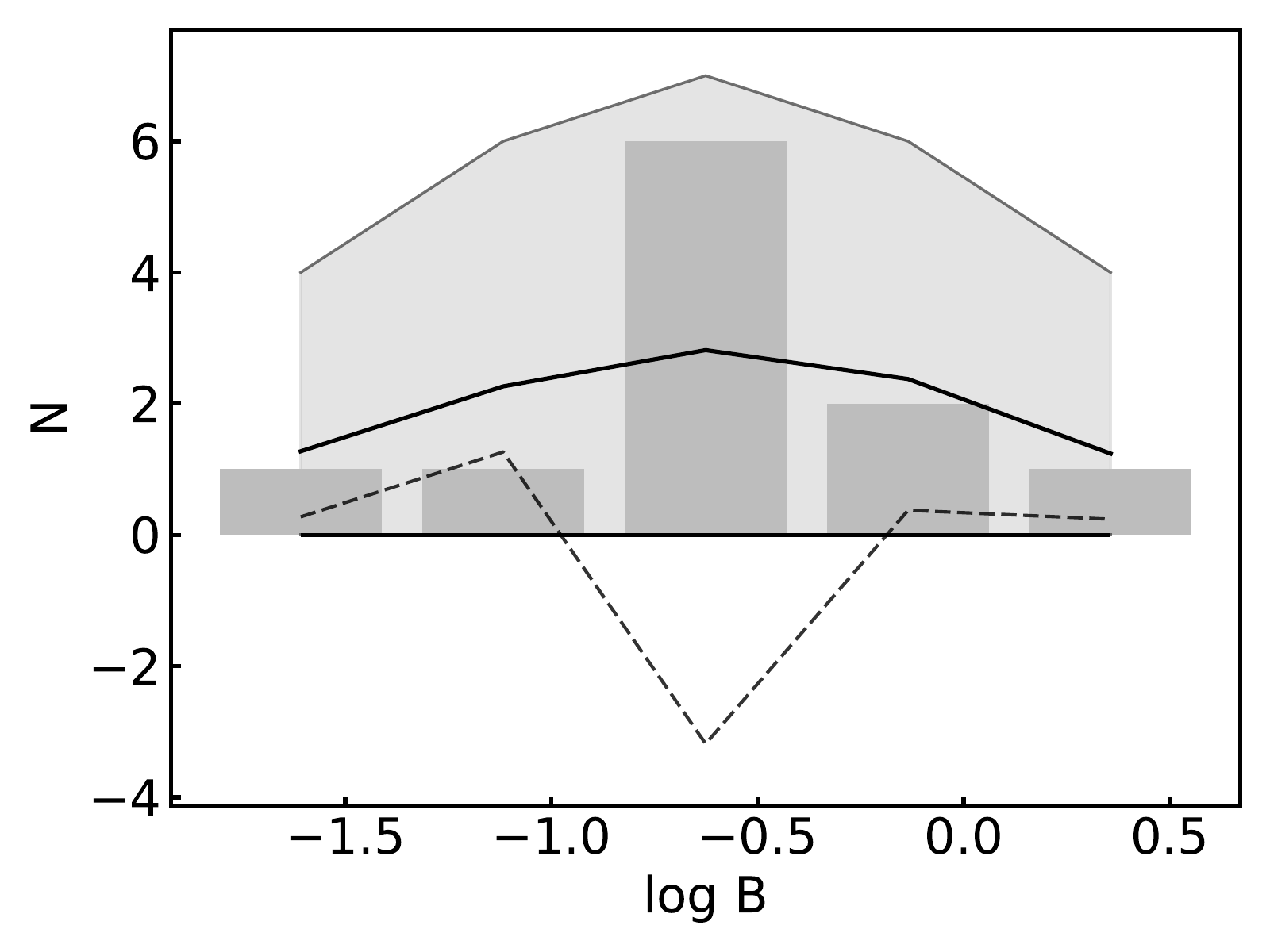}
\caption{Same as in Fig.~\ref{f:Ap/Bp}, but for O-type stars.    
}
\label{f:O}
\end{figure*}

Up to date the number of known O-type magnetic stars is still very small. Consequently, it is 
virtually impossible to make reliable estimates of the parameters describing the corresponding magnetic 
field function from the available data because of large uncertainties of counting statistics. Ne\-vertheless, 
we applied the same procedure as before and found the best-fitting parameters for the models 
$\mathcal{M}_0$ and $\mathcal{M}_1$. A comparison of the model and the empirical magnetic field distributions 
for O stars and for the models $\mathcal{M}_0$ and $\mathcal{M}_1$ is given in Fig.~\ref{f:O}.

We obtained significantly higher values for $\langle{\log\Phi}\rangle$ 
in comparison to the other groups of stars, although the estimated $95$\% confidence intervals for 
the model parameters are turned to be also much wider (Table~\ref{t:fits}).  The null hypothesis 
that the empirical magnetic field distribution of O-type stars is drawn from the same magnetic field 
function as the distributions of BA and OB stars gives the $p$-value of $0.1$, 
which is quite low but still higher than the commonly accepted rejection value of $0.05$.

In order to test further the hypothesis that all of the empirical distributions could be drawn from a single 
magnetic field function, we applied the technique of simultaneous fitting, which was described above in 
Section~\ref{s:parameter_estimation} to all sample of OBA stars including the individual subsamples of BA, 
OB and O stars considered above. The results of the fitting are listed in the last 2 rows of 
Table~\ref{t:fits}. This result supports our hypothesis on the similarity of all of the considered 
empirical distributions. It means that the magnetic field distribution for OBA stars obeys a common 
distribution law.

\subsection{Magnetic Flux distribution}
\label{s:magnetic_flux}

\begin{figure}
\centering
\includegraphics[width=\linewidth]{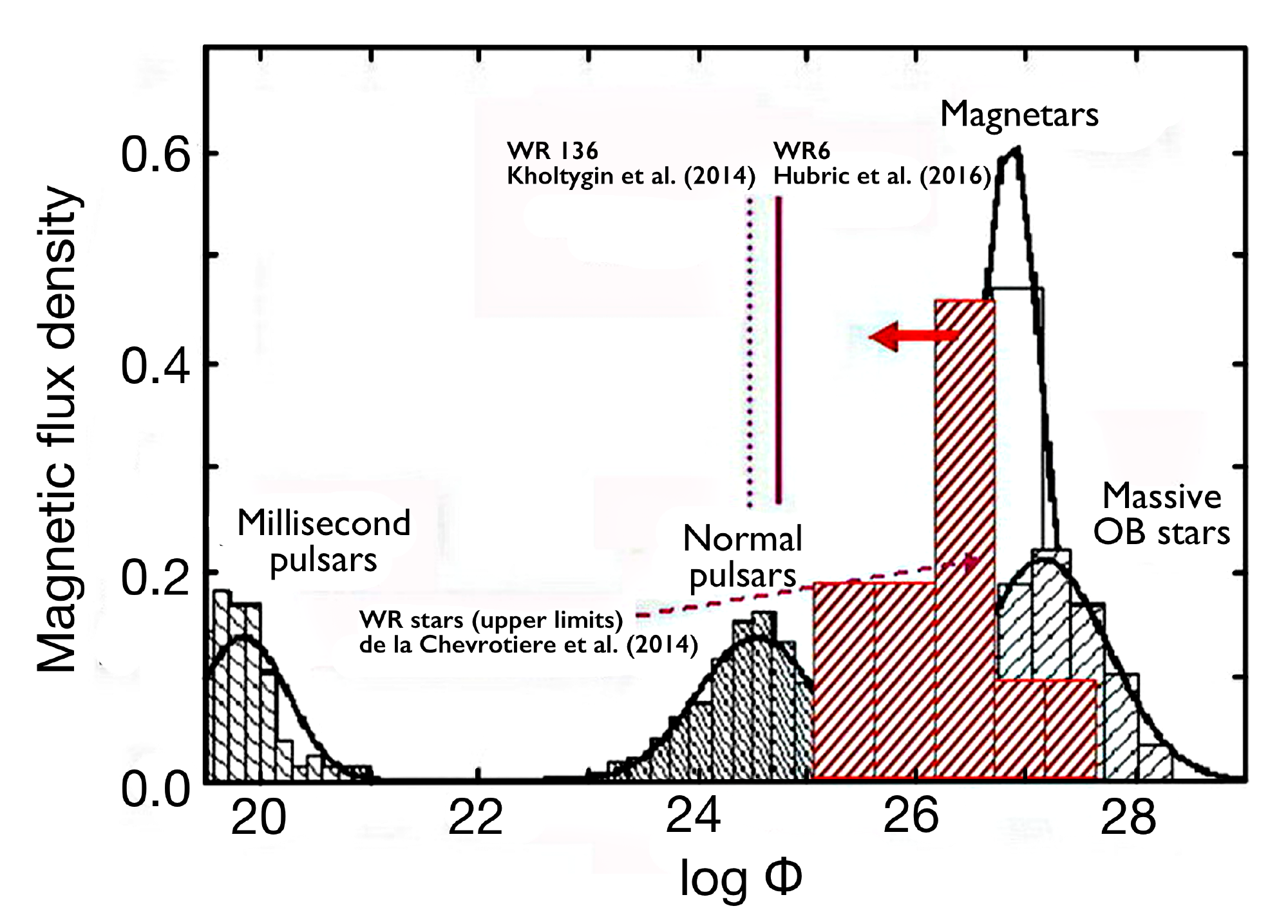}
\caption{The distrubution of magnetic fluxes for massive OB stars, magnetars, and normal and millisecond 
         (thick black lines and dashed histograms). Red dashed historam shows the distribution of the 
         upper limits of the magnetic fluxes for WR stars (using the red arrow we note that the real distribution would be shifted towards smaller values of magnetic flux). The upper limits of the magnetic fluxes
         for WR\,6 and WR\, 136 are shown by solid and dotted red lines, respectively. 
} 
\label{f:MFD}
\end{figure}

In this subsection we outline the evolution of magnetic field of the massive stars after the main sequence. 
In Fig.~\ref{f:MFD} we present the magnetic fluxes distributions for OB stars, Wolf-Rayet (WR) stars and 
different types of the neutron stars (NSs). 
For OB stars we took the value of \emph{rms} magnetic fields from the catalogue by Bychkov et al. (2009). 
The radii of stars were taken either from the original papers for considered stars or from the CADARS 
Catalogue (\cite{CADARS}). 

All known galactic NSs were divided into three groups: normal pulsars, milliseconds pulsars, and magnetars. 
The data for neutron stars are taken from the ATNF Pulsar Catalogue (\cite{Manchester-2005}) and 
\cite{Magnetars}. We put for all NSs the standard radius $R_*=10\,$km. 
The distribution of the magnetic fluxes both for these groups of neutron stars and for OB 
stars are the same what was presented by Igoshev \& Kholtygin (2011) in their Fig.~1.  

We add to their data the distribution of the upper limits of the magnetic fluxes for WR stars. 
The upper limits of magnetic fields for 11 WR stars was taken from the paper \cite{Chevrotiere-2014}. The 
upper limit of the magnetic field for the star WR\,136 is given by Kholtygin et al.\ (\cite{Kholtygin-2011b}) and for 
WR\,6 by \cite{Hubrig-2016}. For all WR stars we use the standard radius $R_* = 5R_{\odot}$.

We can see in Fig.~\ref{f:MFD} that although the width of distributions for NSs are close to that for OB
stars, the mean fluxes are very different. The youngest group of NS (magnetars) has the largest fluxes, 
while the magnetic fluxes of the oldest NS (millisecond pulsars) are on the average smaller by 7 
orders-of-magnitude. The mean fluxes of all remaining NSs (Normal pulsars) have intermediate values. 
We see that the mean fluxes of all types of NSs, excluding magnetars, are much lower than the mean 
fluxes of their progenitors, viz. the massive OB stars.

On the other hand, the upper limits of the magnetic flu\-xes for WR stars are close to those for normal pulsars.
It means that the main dissipation of the magnetic flux during the evolution from the main sequence massive 
OB stars occurs between the ZAMS and the WR stage. At the same time, this can mean that the changes
of the magnetic fluxes during of the supernova explosion are relatively small in agreement with the
hypothesis by \cite{Ferrario-2006}.

\section{Discussion} 
\label{s:discussion} 

\subsection{The ``magnetic desert'' problem}
\label{s:magnetic_desert}

\cite{Lignieres-2014} proposed a scenario whereby the magnetic dichotomy between BA and Vega-like 
magnetism originates from the bifurcation between stable and unstable large scale magnetic configurations 
in differentially rotating stars. This means that the number of BA (and possibly O) stars with  
the \emph{rms}-magnetic fields in interval between $\sim$1\,G and $\sim$300\,G is small if not 
negligible.  

\begin{figure}
\centering
\includegraphics[width=\linewidth]{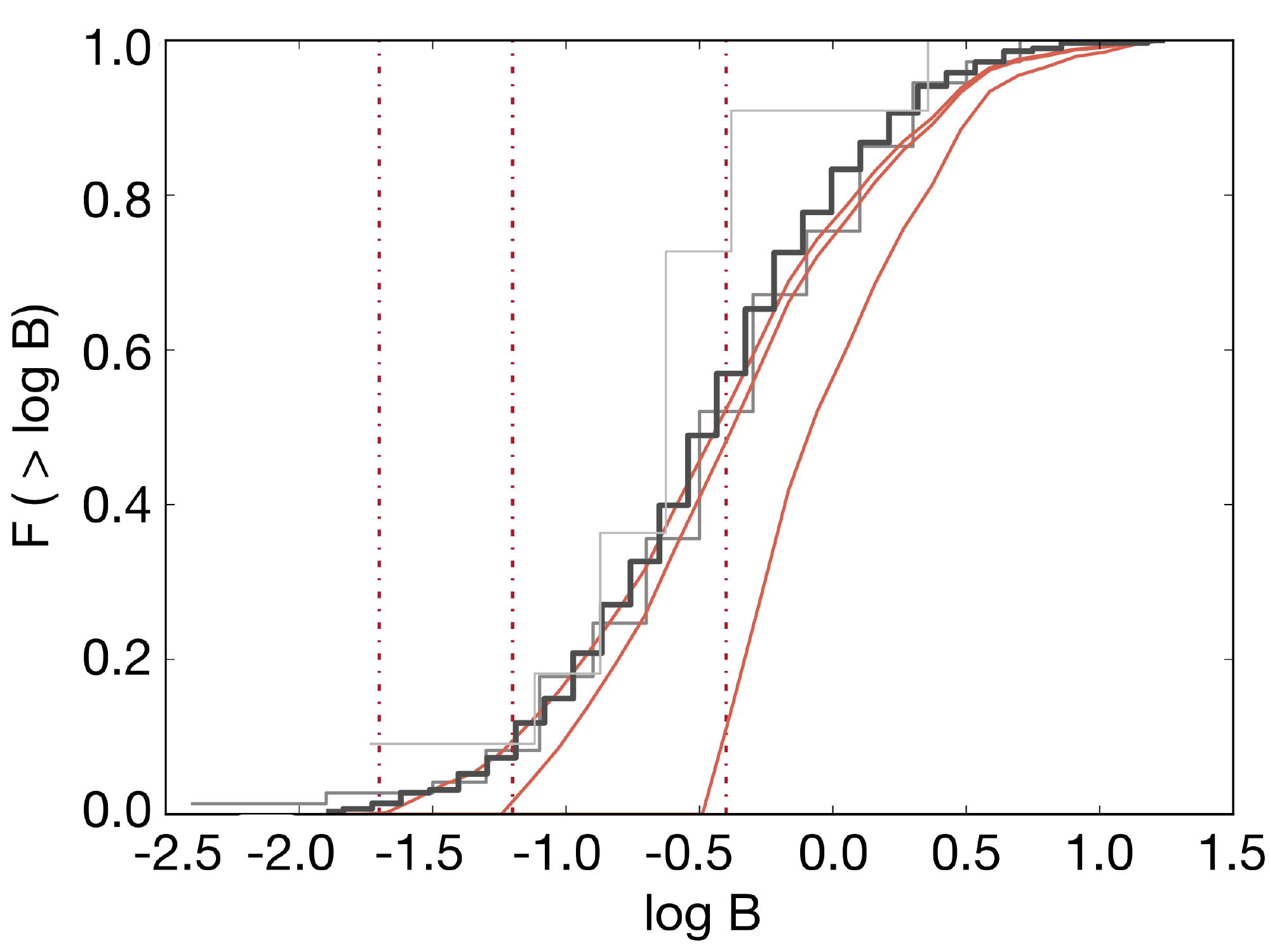}
\caption{Cumulative distributions and magnetic thresholds for data from different papers (from left to right: 
\cite{Wade:2015vq}; \cite{Auriere:2007dw}; \cite{Kholtygin-2010a}). 
}
\label{f:cdf}
\end{figure}

At the same time the empirical magnetic field functions obtained by us for OB and BA stars reveal that
the same regular shape can be fitted by a lognormal distribution. There are no visible peculiarities that 
might indicate the existence of a threshold magnetic field reported for BA and early B-type stars 
by Auri\'ere et al. (2007) and by Kholtygin et al. (2010a). The cumulative distributions produced by 
truncated magnetic field functions are too steep in comparison to the empirical ones 
(see Fig.~\ref{f:cdf}).

The same issue also has been discussed by Fossati et al. (2015a), who reported the detection of weak magnetic 
fields (below $100$~G) in two early B-type stars: $\beta$~CMa (HD 44743) and $\epsilon$~CMa (HD 52089). 
The estimated upper limit for a dipolar field strength of the latter star is $B_d \gtrsim 13\,$G, which is 
far beyond the threshold value $B_d \approx 300\,$G obtained by Auri\'ere et al. (2007),
and $B_d \approx 1500\,$G obtained by 
Kholtygin et al. (2010a). Based on the presented considerations, we also conclude
that there is no evidence for any `magnetic threshold' at least for BA and late B-type 
stars.

On the other hand, Wade et al. (2015) reported preliminary results on a study where a modeling approach by 
\cite{Petit:2012cg} has been applied to infer upper limits on dipole magnetic fields in a
sample of O-type stars from the MiMeS survey. The results of \cite{Wade:2015vq} imply 
upper limits on surface dipoles of about 40~G at 50\% confidence, and $105$~G at $80$\% confidence. The later values are more consistent with the empirical distributions 
(Fig.~\ref{f:cdf}). Therefore, we cannot rule out the possibility that the critical value for a magnetic 
field may exist, but this value should be much lower than proposed earlier.

\subsection{Constrains on dissipation of stellar magnetic fields}
\label{s:constrains_on_dissipation}

An important result of our analysis of the empirical magnetic fields distributions is that the time scale 
of the magnetic fields dissipation should be at least comparable with the lifetime of a star on the main 
sequence. Using the technique of simultaneous fitting we found that the dissipation parameter has to be greater
 than $0.5$ to be consistent with the empirical distributions. This result is in a good agreement with the 
value obtained by Kholtygin et al. (2010b). However, we cannot estimate the dissipation parameter 
$\tau_\mathrm{d}$ more precisely with the current sample of magnetic stars. It means that models with 
moderate dissipation and without the dissipation all are equally consistent with the empirical data.

We would also like to note, that according to Landstreet et al. (2008), the mean magnetic fields of early 
A- and late B-type stars decrease by a factor of $3$--$4$ in the range $\tau \in (0.0$--$0.2)$ and then 
remain almost constant. The dissipation parameter in that case would be lying within the range of 
$0.1$--$0.2$and, according to our simulations, would produce a highly asymmetrical shape of the magnetic 
field function, inconsistent with the empirical distributions. We believe that this discrepancy may be 
explained by a limited size of the magnetic stars sample used in their analysis.

\subsection{Intrinsic magnetic field function for early-type stars}
\label{s:intrinsic_MFF}

As we discussed earlier, the empirical distributions are very similar in their appearance. Moreover, it is 
possible to successfully describe all three of our samples of magnetic stars with a single model. Thus, 
we consider the possibility that magnetic properties of early type massive stars may be explained by a 
common magnetic field function {\it that would be the same} in a very wide range of masses 
($\gtrsim 2 M_{\odot}$).

Another strong evidence supporting this hypothesis is that the occurrence of early type magnetic stars 
is remarkably constant and does not depend on a spectral type or mass of a star. The incidence of magnetic
 BA stars -- which is slightly less than ten per cent -- has been known for some time 
(\cite{Wolff:1968jt}; \cite{Power:2008wd}). 

Recently similar magnetic field detection rates were reported for early 
B- and O-type stars by the BOB  (\cite{Scholler-2017}) and MiMeS collaborations \cite{Wade:2016es}. It is 
certainly became more evident now that magnetic stars across the upper main sequence share some common 
phy\-sics related to the phenomena of stellar magnetism (see also \cite{Wade:2014cj}, 2015). 
Gi\-ven the fact that the incidence of magnetic Herbig Ae/Be stars is also estimated as 7\% 
(\cite{Wade:2011wt}), we come to the conclusion that magnetic properties of massive stars are 
probably determined by physical conditions at early stages of pre-main-sequence evolution, perhaps 
even during protostar formation.

\section{Conclusions}
\label{s.Concl}

\noindent
Our investigations of the measured magnetic fields and their evolution for an ensemble of galactic OBA stars show that 
the distribution of these magnetic field for all mentioned spectral types can be described by a lognormal law. 
We develop the population synthesis code to model the magnetic field evolution for OBA stars.

After the analysis of the empirical magnetic field distributions using our model we came to the following conclusions:

\begin{itemize}

\item  The empirical magnetic field distribution for BA stars can be fitted by the lognormal distribution with 
       the mean $\langle{\log{B}}\rangle \approx -0.5$ and the standard deviation  $\sigma = 0.5$.

\item  Our model can be used to reproduce the empirical magnetic field and net magnetic flux distributions 
       for OBA stars supposing that 
       the net magnetic flux distributions at ZAMS is lognormal with parameters 
       $\langle\log\Phi\rangle \approx 26.5$ for model without magnetic field 
       dissipation and for a model  with dissipation $\langle\log\Phi\rangle \approx 26.9$.

\item  The dissipation parameter $\tau_\mathrm{d} \gtrsim 0.5$ in an accordance with an estimation 
       by \cite{Kholtygin-2010b}.
 
\item  Our modeling shows that there is no {\it magnetic desert} in the distribution of the magnetic field of 
       OBA star suggested in the past by some authors (e.g., \cite{Lignieres-2014}).

\end{itemize}
       
\acknowledgements
    ASM, AFK, SF, GGV and GAC thank the RFBR grant 16-02-00604~A for the support.
    GGV acknowledges the Russian Foundation for Basic Research (RFBR grant N15-02-05183~A)


\end{document}